\documentclass[preprint,3p,twocolumn]{elsarticle}

\usepackage{amssymb}

\usepackage{amsmath}

\usepackage{upgreek}

\usepackage{lineno}

\usepackage{color}

\journal{Science Bulletin}

\begin{document}

\begin{frontmatter}

\title{Giant anomalous Hall and Nernst effects in a heavy fermion ferromagnet}

\author[aff1,aff6]{Longfei Li\corref{cor1}}

\author[aff1]{Shuyue Guan\corref{cor1}}

\author[aff2]{Shengwei Chi\corref{cor1}}

\author[aff3]{Jianzhou Zhao}

\author[aff1]{Jiawei Li}

\author[aff1]{Xinxuan Lin}

\author[aff2]{Gang Xu}
\ead{gangxu@hust.edu.cn}

\author[aff1,aff4,aff5]{Shuang Jia}
\ead{gwljiashuang@pku.edu.cn}

\address[aff1]{International Center for Quantum Materials, School of Physics, Peking University, Beijing 100871, China}
\address[aff6]{Department of Physics, Southern University of Science and Technology, Shenzhen 518055, China}
\address[aff2]{Wuhan National High Magnetic Field Center and School of Physics, Huazhong University of Science and Technology, Wuhan 430074, China}
\address[aff3]{Department of Physics, School of Science, Tianjin University, Tianjin 300354, China}
\address[aff4]{Interdisciplinary Institute of Light-Element Quantum Materials and Research Center for Light-Element Advanced Materials, Peking University, Beijing 100871, China}
\address[aff5]{Hefei National Laboratory, Hefei 230088, China}

\cortext[cor1]{These authors contribute equally.}

\begin{abstract}
The anomalous Hall and Nernst effects refer to the perpendicular voltage drop generated by a magnetic material's magnetization in response to an applied current and temperature gradient.
These effects can be harnessed to determine the Berry curvature and hold potential for future applications in electronic devices and thermoelectric energy conversion.	
We investigate the anomalous Hall and Nernst effects in the heavy-fermion ferromagnet CeCrGe$_3$ and its non-4f analog ferromagnet LaCrGe$_3$. 
We find that CeCrGe$_3$ exhibits a giant anomalous Hall angle and an anomalous Nernst coefficient, reaching values as high as 33\% and $\sim 10~\upmu\mathrm{V\,K}^{-1}$, respectively, among the largest reported for topological magnets. 
Based on electronic band-structure calculations, we identify a series of topological flat bands carrying strong Berry curvature with a pronounced Ce 4f orbital character in CeCrGe$_3$, which are absent in LaCrGe$_3$, highlighting the crucial role of Kondo flat bands in generating large anomalous transport responses. 
Furthermore, we identify a breakdown of the anomalous Hall scaling relation and the nonlinear anomalous Mott relation, which we attribute to the break of the topological Kondo flat bands at finite temperatures.
\end{abstract}

\begin{graphicalabstract} 
The anomalous Hall angle and anomalous Nernst coefficient reach values as high as 
$33$\% and $\sim 10~\upmu\mathrm{V\,K}^{-1}$ in the ferromagnetic heavy-fermion CeCrGe$_3$, respectively, placing the compound among the topological magnets with the largest reported anomalous responses.
These giant anomalous effects originate from the large Berry curvature in the topological Kondo flat bands near the Fermi energy.
Our findings pave the way for further exploration of giant topological responses driven by electron correlation.
\begin{figure}[!htp] 
	\centering 
	\includegraphics[width=1\linewidth]{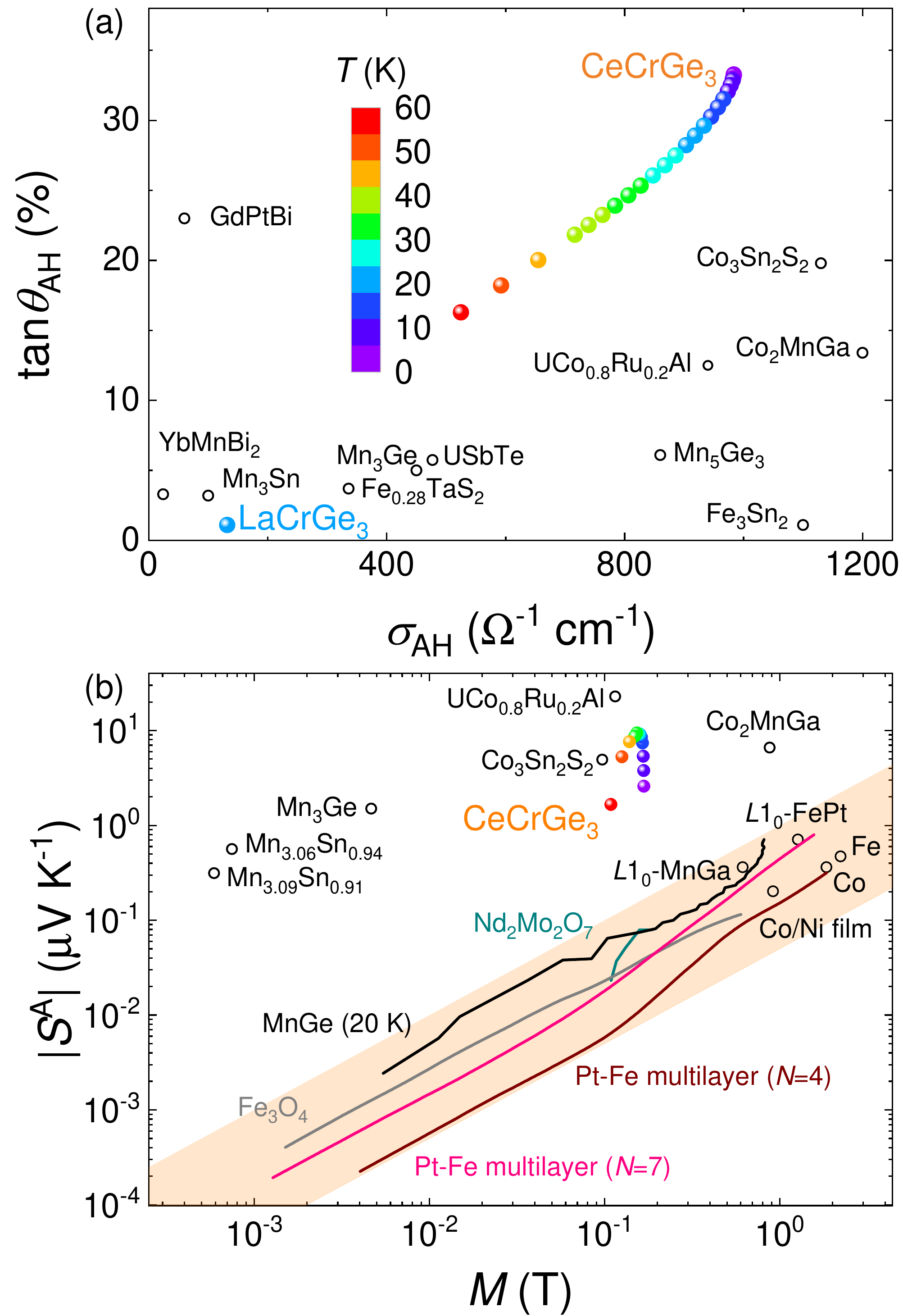} 
\end{figure}
\end{graphicalabstract}


\end{frontmatter}

\section{Introduction}

Investigating the connection between electron correlation and topology is a pivotal goal in current condensed matter physics research \cite{Pesin2010, Tokura2022}.
Strong correlations are necessary for the emergence of various topological phases, such as the fractional quantum Hall effect \cite{RevModPhys.71.S298}, topological Kondo insulators \cite{PhysRevLett.104.106408}, and the anomalous fractional quantum Hall effect \cite{Park2023} that was recently discovered in experiments. Heavy-fermion (HF) systems are a prototypical class of strongly correlated electron materials, in which Kondo entanglement between localized f-electron moments and conduction electrons gives rise to strong hybridization\cite{hewson_1993,doi:https://doi.org/10.1002/9780470022184.hmm105}. 
At low temperatures, the overlap of electronic states in HF compounds leads to a coherent Kondo state with a drastically narrowed bandwidth and an energy scale orders of magnitude smaller than that of ordinary metals. 
The extremely low Fermi energy ($E_{\mathrm{F}}$) and large effective mass ($m^{\ast}$) of the HF systems allow for the investigation of topological states with strong correlation. Recent studies have even suggested the existence of various topological states in HF compounds, such as Weyl-Kondo semimetals and topological semimetals without quasiparticles~\cite{PhysRevB.101.075138,doi:10.1073/pnas.1715851115,doi:10.1073/pnas.2013386118,hu2022topological,cheng2023observation}.

Two of the most significant phenomena in topological magnets are the anomalous Hall effect (AHE) and anomalous Nernst effect (ANE), which manifest as transverse electric and thermoelectric responses in magnetic materials even in the absence of an external magnetic field~\cite{RevModPhys.82.1539}.
Topological magnets can generate strong Berry curvature (BC), leading to anomalous effects much stronger than those in conventional magnets \cite{Ikhlas2017,Sakai2018,https://doi.org/10.1002/adma.201806622,Sakai2020,10.1063/5.0005481,Pan2022,doi:10.1126/sciadv.abk1480,https://doi.org/10.1002/advs.202207121}, with important applications in dissipationless transport~\cite{RevModPhys.82.1539} and transverse thermoelectric devices~\cite{Chen2024Topological,Chen2025Design}. However, a major bottleneck for thermoelectric applications is that the anomalous Nernst coefficient ($S^{\mathrm{A}}$) of ordinary metals is typically only on the order of 1~$\upmu$V K$^{-1}$~\cite{Ramos2014Fe3O4,Hasegawa2015Fe, Chuang2017FeCoNi}.
The weak-correlation topological magnets exhibit relatively large anomalous Nernst effects, such as Co$_2$MnGa and Co$_3$Sn$_2$S$_2$, in which the $S^{\mathrm{A}}$ values were reported in the range of $3$--$6~\upmu\mathrm{V\,K^{-1}}$~\cite{Sakai2018,https://doi.org/10.1002/adma.201806622,PhysRevX.9.041061,Guin2019,PhysRevB.101.180404,PhysRevMaterials.4.024202}. 
In contrast, it was reported that the strongly correlated magnet UCo$_{0.8}$Ru$_{0.2}$Al owns a much greater $S^{\mathrm{A}}$ of $23~\upmu\mathrm{V\,K^{-1}}$.
It is intriguing to ask whether the anomalous Nernst response in the HF magnets in general exceeds that in weakly correlated magnets.

Here, we report giant anomalous Hall and Nernst effects in CeCrGe$_3$, an HF ferromagnet with a high Curie temperature ($T_{\mathrm{C}} = 65$~K), which is unusually high for Ce-based compounds~\cite{Li2025Extrinsic,Wirth2016exploring}. Thermodynamic and transport measurements confirm its HF nature and reveal an anomalous Nernst coefficient as large as $-S^\mathrm{A}_{zx} = 9.5~\upmu$V K$^{-1}$ at 35~K, among the largest recorded ones~\cite{Ikhlas2017, Sakai2018, PhysRevMaterials.4.024202, Chen2021, doi:10.1126/sciadv.abf1467}. This observation provides guidance for discovering topological thermoelectric materials and advancing energy conversion technologies.

Additionally, we observed a giant AHE, characterized as significant values of anomalous Hall conductivity (${\sigma^{\mathrm{A}}_{zx}}$) and anomalous Hall angle ($\tan \theta_{\mathrm{AH}}$) which is defined as the ratio of the anomalous and longitudinal conductivities, as large as 983 $\Omega$$^{-1}$ cm$^{-1}$ and 33\%, respectively, both among the largest values of various topological magnets~\cite{Ikhlas2017,Sakai2018,PhysRevMaterials.4.024202,Chen2021,doi:10.1126/sciadv.abf1467}.  
Through electronic structure calculation, we recognized a strong hybridization between the Ce 4f and Cr 3d electrons, leading to the formation of a Kondo flat band near the Fermi level.
More importantly, our calculation revealed a strong BC in a narrow energy window, which aligns with the measured ${\sigma^{\mathrm{A}}_{zx}}$ and anomalous Nernst conductivity ($\alpha^{A}_{zx}$).
In comparison with the weak AHE in the non-4f conventional magnet LaCrGe$_3$, we emphasize the crucial role of the coherent Kondo effect in CeCrGe$_3$, which provides a large density of states (DOS) and strong BC that underlie its giant anomalous effects.
Moreover, we observe a breakdown of the AHE scaling relation and a nonlinear anomalous Mott relation at elevated temperatures due to the Kondo band instability.
Our finding reveals a mechanism for generating significant BC in a topologically flat band residing in a scarce example of a ferromagnetic (FM) Kondo lattice, which paves the way for further exploration of giant topological responses driven by electron correlation.

\section{Materials and methods}

High-quality single crystals of LaCrGe$_3$ and CeCrGe$_3$ were grown by the self-flux method \cite{PhysRevB.88.094405}. The raw materials of rare earth, chromium, and germanium ingots were mixed in a molar ratio of 15:15:85 and placed in a dry tantalum crucible, which was then sealed in a quartz ampoule under vacuum. The ampoule was placed in a furnace, slowly heated up to 1100~°C over 7 h, and held for 10 h. It was then cooled down to the centrifugation temperature ($\sim 850$~°C), where the excess self-flux was centrifuged. 
The single crystals of CeCrGe$_3$ and LaCrGe$_3$ are as large as $4\times2\times1$~$\mathrm{mm}^3$.
The crystal structure was examined by powder X-ray diffraction (PXRD) using a Rigaku MiniFlex 600 diffractometer with Cu K$_{\alpha}$ radiation.

Magnetic properties were measured using the Quantum Design Magnetic Property Measurement System (MPMS 3), with the magnetic field applied along the directions of the crystallographic $c$ axis and the $ab$ plane, respectively.
Transport properties were measured with the Quantum Design Physical Property Measurement System (PPMS) and the Oxford TeslatronPT system.
Longitudinal and Hall resistivity were measured using the standard four-probe method, and the Lake Shore Model 372 AC resistance bridge was used to collect the signals. Note that we define the crystallographic $a$ axis, the direction perpendicular to the $a$ and $c$ axes, and the $c$ axis as the $x$, $y$, and $z$ directions, respectively. We adopt the convention $E_i = \rho_{ij} J_j$, such that $\rho_{xz}$ is positive for hole carriers when the current is applied along the $z$ axis, and the magnetic field is along the $y$ axis.
Thermoelectric measurement was taken with a one-heater-three-thermometer setup using Cernox thermometers to monitor both the longitudinal and transverse temperature gradient simultaneously. The thermoelectric voltages were amplified with the EM DC Nanovolt Amplifier Model A10 and then collected by the Keithley Nanovoltmeter Model 2182A. Note that we use the convention $E_{i}=S _{ij} \partial_{j} T$.
Detailed descriptions of the transport measurement setups and the procedures for obtaining the transport coefficients are provided in the Supplementary Material.

In order to capture the electronic correlation effect, fully charge self-consistent density functional theory plus dynamical mean-field theory (DFT + DMFT) calculations were performed.
Hubbard $U$ = 6.0 eV, and Hund's coupling $J$ =0.7 eV of Ce 4f electrons were used.
We renormalized the hopping and onsite energy parameters of the tight-binding Hamiltonian to reproduce the DFT + DMFT calculated band structures.  Based on the DFT + DMFT calculations, the hopping renormalization factors for the 4f orbitals of Ce atoms are 0.06853 and 0.32991 for $J_z = 5/2$ states and $ J_z = 7/2 $ states, respectively.
The onsite energies are -0.207 eV for the $J_z = 5/2$ states and 0.893 eV for the $J_z = 7/2$ states.

Because direct determination of the BC is not feasible in DFT+DMFT calculations, the first-principles calculations based on DFT were implemented in the Vienna ab initio simulation package \cite{KRESSE199615, PhysRevB.54.11169} with the projector augmented wave method \cite{PhysRevB.50.17953} to obtain anomalous Hall conductivity ($\sigma^{\mathrm{A}}_{\alpha\beta}$) and anomalous thermoelectric conductivity ($\alpha^{\mathrm{A}}_{\alpha\beta}$) . The cutoff energy for wave function expansion is set as 400 eV with 11×11×11 $k$ meshes. The LDA type of exchange-correlation potential \cite{PhysRevB.23.5048} is utilized in our calculations. 
A Wannier tight-binding Hamiltonian consisting of 4f and 5d orbitals of Ce atoms, 3d orbitals of Cr atoms, and 4p orbitals of Ge atoms was constructed using the Wannier90 package \cite{MOSTOFI20142309}. Our DFT band structure exhibits results that are highly consistent with those obtained from DMFT calculations, as shown in Fig.~S11 (online) of the Supplementary material.

We used WannierTools \cite{WU2018405} to calculate the $\sigma^{\mathrm{A}}_{\alpha\beta}$ value and find Weyl points. The $\sigma^{\mathrm{A}}_{\alpha\beta}$ and $\alpha^{\mathrm{A}}_{\alpha\beta}$ were obtained by integrating the BC according to the formula \cite{PhysRevB.74.195118,BerryXiao2006}:

\begin{linenomath*}
	\begin{equation}
		\sigma^{\mathrm{A}}_{\alpha\beta}=-\frac{e^2}{\hbar}\int_{\mathrm{BZ}}\frac{\mathrm{d} \boldsymbol{k}}{(2\pi)^3}\sum_{n}f(\varepsilon_{n\boldsymbol{k}})\Omega_{n,\alpha\beta}(\boldsymbol{k}),
	\end{equation} 
\end{linenomath*}

\begin{linenomath*}
	\begin{equation}
		\alpha^{\mathrm{A}}_{\alpha\beta}=\frac{ek_{\mathrm{B}}}{\hbar}\int_{\mathrm{BZ}}\frac{\mathrm{d} \boldsymbol{k}}{(2\pi)^3}\sum_{n}s(\varepsilon_{n\boldsymbol{k}})\Omega_{n,\alpha\beta}(\boldsymbol{k}).
	\end{equation} 
\end{linenomath*}

The BC of band $n$ with momentum $\boldsymbol{k}$ is
\begin{linenomath*}
	\begin{equation}\Omega_{n,\alpha \beta}(\boldsymbol{k})= -2 \hbar^2 \mathrm{Im} \sum_{m(\neq n)} \frac{\langle n\boldsymbol{k}|v_{\alpha}| m\boldsymbol{k} \rangle \langle m\boldsymbol{k}|v_{\beta}| n\boldsymbol{k} \rangle}{(\varepsilon_{n\boldsymbol{k}}-\varepsilon_{m\boldsymbol{k}})^2},
	\end{equation} 
\end{linenomath*}
where $v_{\alpha}$ is the velocity operator along the $\alpha$ direction, and $| n\boldsymbol{k} \rangle$ is the Bloch state with band index $n$ and momentum $\boldsymbol{k}$. $f(\varepsilon)=(e^{(\varepsilon-\mu)/k_{\mathrm{B}}T}+1)^{-1} $ is the Fermi distribution function, and $s(\varepsilon)$ is given as $-f(\varepsilon)\ln f(\varepsilon)-(1-f(\varepsilon))\ln(1-f(\varepsilon))$.

\begin{figure*}[!htp] 
	\centering 
	\includegraphics[width=1\textwidth]{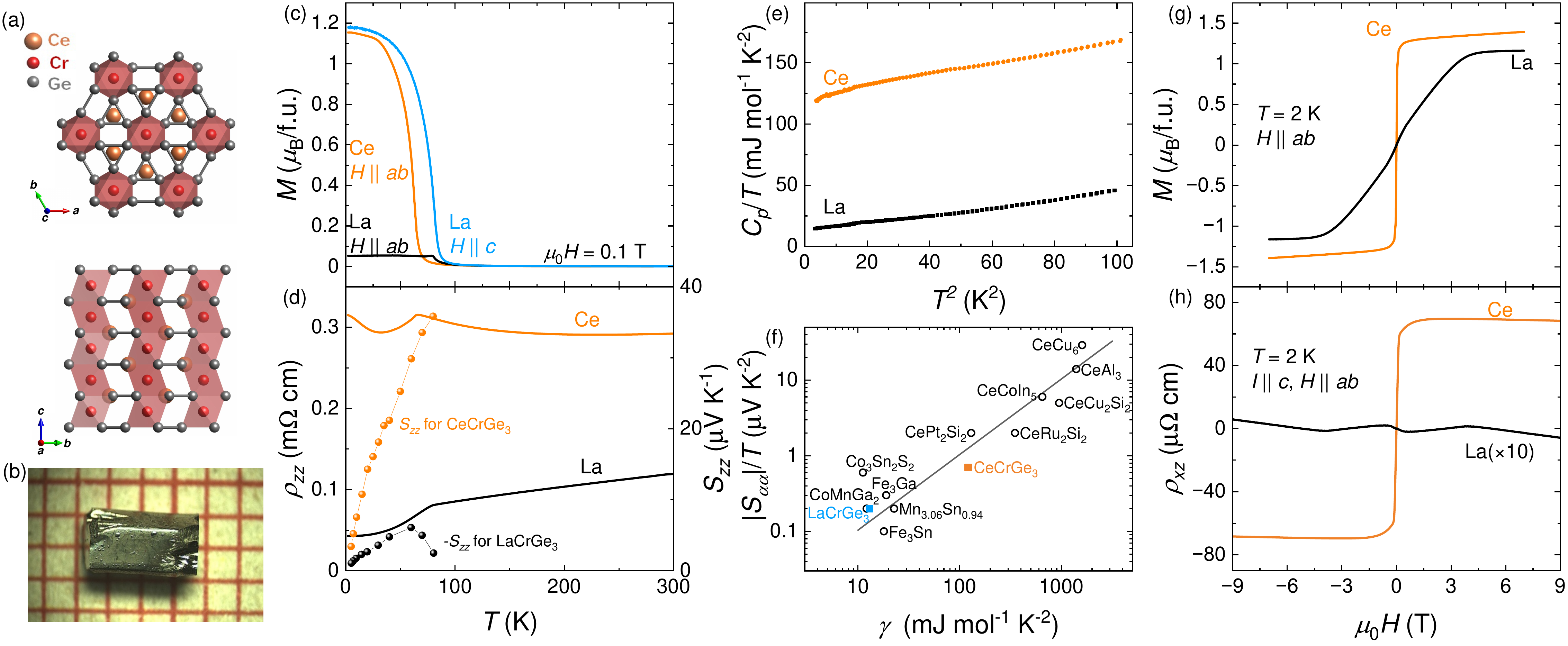} 
	\caption{Crystal structure and physical properties of CeCrGe$_3$ and LaCrGe$_3$.
		(a) Crystallographic structure of the hexagonal perovskite-type CeCrGe$_3$.
		(b) Photo of a single crystal of CeCrGe$_3$.
		(c) The temperature dependence of the magnetization in an external magnetic field of 0.1 T applied in the crystallographic $ab$ plane (black and orange curves) and $c$ axis (the blue curve). 
		(d) The temperature dependence of the longitudinal electric resistivity $\rho_{zz}(T)$, and Seebeck coefficient  $S_{zz}(T)$  of CeCrGe$_3$ and LaCrGe$_3$ in zero field.  
		(e) Low-temperature specific heat of LaCrGe$_3$ and CeCrGe$_3$, presented as $C_p/T$ versus $T^2$.
		(f) Plot of $|S_{\alpha\alpha}|/T$ versus $\gamma$ for representative Ce-based HF compounds and topological magnetic materials (the Supplementary material, Table S1 online.). 
		(g, h) The magnetic-field dependence of the magnetization and the Hall resistivity at 2 K, with the field applied in the $ab$ plane and the current applied along the $c$ axis.} 
	\label{Figure1} 
\end{figure*}

\section{Experimental results}

\begin{figure*}[!htp] 
	\centering 
	\includegraphics[width=0.9\textwidth]{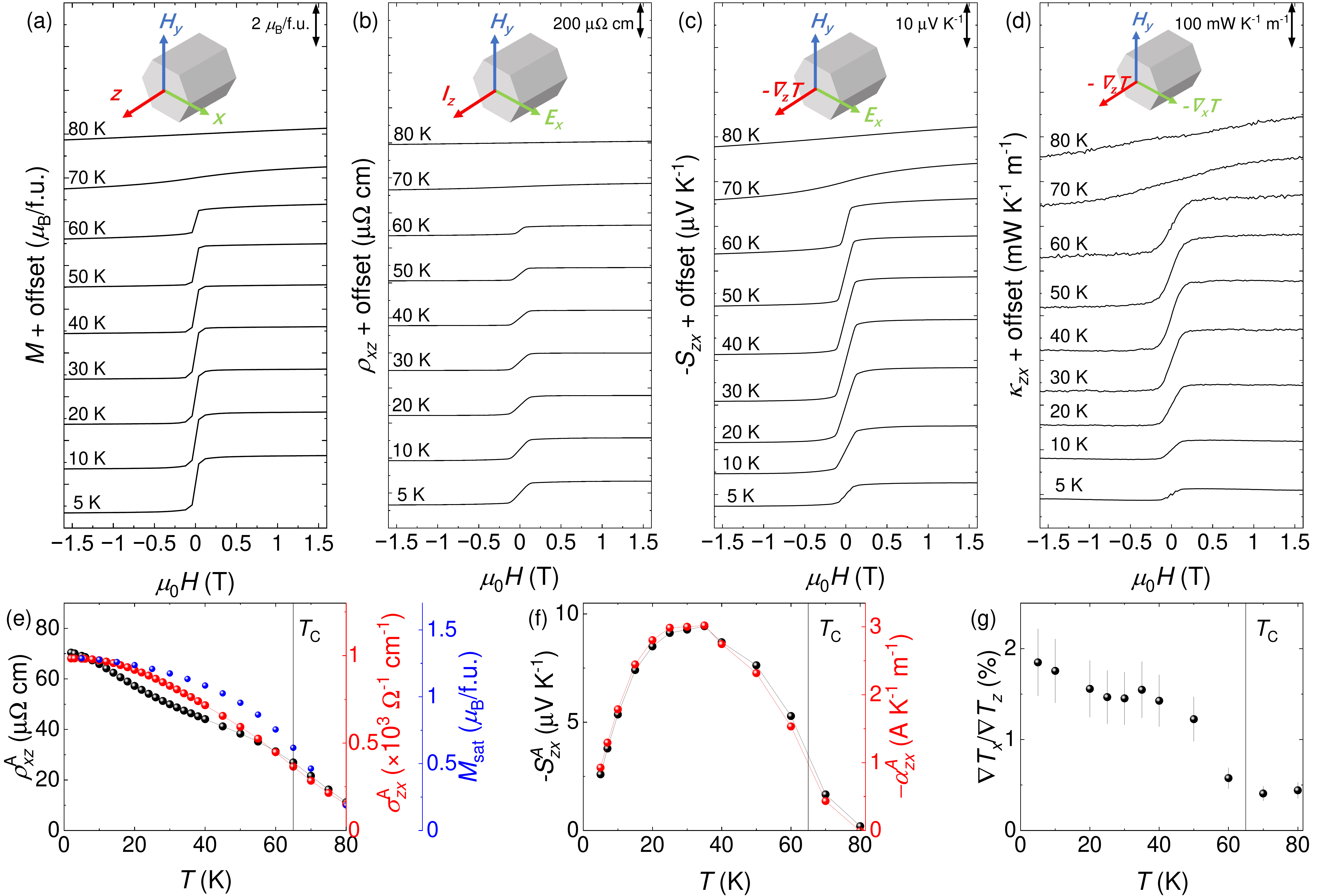} 
	\caption{Anomalous transport in CeCrGe$_3$.
		(a-d) The field dependence of the magnetization $M$, the Hall resistivity $\rho_{xz}(H)$, the Nernst thermopower $-{S_{zx}}(H)$, and the thermal Hall conductivity $\kappa_{zx}(H)$  at representative temperatures. Curves are shifted vertically for clarity. 
		(e) The temperature dependence of $\rho^{\mathrm{A}}_{xz}$  (black, left axis), $\sigma^{\mathrm{A}}_{zx}$ (red, right axis), and saturated magnetization (blue, right axis).
		(f) The temperature dependence of $-S^{\mathrm{A}}_{zx}$ (black, left axis) and $-\alpha^{\mathrm{A}}_{zx}$ (red, right axis), respectively.
		(g) The temperature dependence of the anomalous thermal Hall angle $\nabla T_x/\nabla T_z$.}
	\label{Figure2} 
\end{figure*}

CeCrGe$_3$ and LaCrGe$_3$ crystallize in the hexagonal perovskite BaNiO$_3$-type structure with space group $P6_3/mmc$. This structure consists of chains of face-sharing Cr-centered Ge octahedra aligned along the crystallographic $c$ axis and separated by Ce/La atoms~\cite{doi:10.1021/cm071276+, PhysRevMaterials.5.114406}, as illustrated in Fig.~\ref{Figure1}a. Figure~\ref{Figure1}b shows a photograph of a single crystal of CeCrGe$_3$.

The Ge atoms on the nearby octahedra's apex are strongly linked, forming a breathing kagome lattice in the $ab$ plane.
Figure \ref{Figure1}c illustrates the temperature dependence of the magnetization in an external field of $0.1$ T for the two compounds. Our findings of their magnetization are in line with previous reports \cite{cadogan2013, Das_2015, PhysRevB.94.174415, SINGH2019411701}. While LaCrGe$_3$ with a 4$f^0$ electronic configuration exhibits FM ordering below the $T_{\mathrm{C}}$ = 78 K with an easy-$c$-axis anisotropy, CeCrGe$_3$ with a 4$f^1$ electronic configuration has a close $T_{\mathrm{C}}$ = 65 K with an easy-$ab$-plane anisotropy.

In contrast, the electric resistivity $\rho_{zz}(T)$ of the two compounds displays distinctive profiles (see Fig. \ref{Figure1}d), revealing a significant role of the Ce 4f electrons on the electron scattering.  
While the $\rho_{zz}(T)$ of LaCrGe$_3$ has a metallic profile, that of CeCrGe$_3$ shows a $-\ln T$ dependence with decreasing temperature down to $T_{\mathrm{C}}$, a characteristic feature of incoherent Kondo scattering \cite{Das_2014,doi:10.1073/pnas.2001778117}.
However, unlike the $\rho(T)$ curves of ordinary Kondo lattices, which typically exhibit a broad coherence peak, the resistivity of CeCrGe$_3$ shows a sharp drop from $T_{\mathrm{C}}$ down to 32~K. This behavior likely originates from the suppression of spin-disorder scattering below $T_{\mathrm{C}}$. On the other hand, it may also suggest the formation of Kondo coherence as long as the magnetic moments of Cr atoms are ordered.
Below 32 K, $\rho_{zz}(T)$ increases again.
A similar profile of the $\rho(T)$ curve was observed in USbTe \cite{USbTe2023Ran} and explained as a result of residual incoherent single-ion Kondo scattering along with coherent Kondo scattering.

Figure \ref{Figure1}e demonstrates that the specific heat of both LaCrGe$_3$ and CeCrGe$_3$ is approximated by $C(T) = \gamma T + \beta T^3$ at low temperatures, where $\gamma T$ is the electronic specific heat and $\beta T^3$ is the lattice contribution.
While the $\beta$ values are close for the two compounds, the value of $\gamma$ is $121\ \mathrm{mJ\ mol^{-1}\ K^{-2}}$ for CeCrGe$_3$, which is one order of magnitude greater than that for LaCrGe$_3$ ($13\ \mathrm{mJ\ mol^{-1}\ K^{-2}}$).
The large $\gamma$ value is one of the characteristics of HF systems whose coherence temperature can be roughly estimated, based on a simple argument of magnetic entropy, as $T_{\mathrm{coh}}={R\ \ln 2}/(\gamma_{\mathrm{Ce}}-\gamma_{\mathrm{La}}) \simeq 50$ K for a doublet ground state, where $R$ is the gas constant \cite{FISK1988637}.
We note $T_{\mathrm{coh}}$ is lower than its $T_{\mathrm{C}}$, consistent with the observation that the coherence peak is absent in the $\rho_{zz}(T)$ curve. 

Another characteristic feature of the HF system is the enhanced Seebeck coefficient at low temperatures.
As shown in Fig. \ref{Figure1}d, the Seebeck coefficients $S_{zz}$ for the two compounds show a linear temperature dependence below 40 K.
The large slope of the $S_{zz}$($T$) of CeCrGe$_3$ at low temperatures correlates with the large $\gamma$, as observed in other Ce-based HF systems (Fig. \ref{Figure1}f) \cite{Behnia_2009,Kamran_Behnia_2004}.
This scaling relation can be explained by a single-band model, where both \( S_{zz}/T \) and \( \gamma \) are proportional to the DOS at $E_{\mathrm{F}}$.
All the above properties demonstrate that CeCrGe$_3$ is an HF ferromagnet which undergoes magnetic ordering at a temperature above its $T_{\mathrm{coh}}$, whereas LaCrGe$_3$ is a regular ferromagnet. 

We compare the magnetic field dependence of the magnetization and Hall resistivity $\rho_{xz}$ for LaCrGe$_3$ and CeCrGe$_3$ at 2 K in Fig. \ref{Figure1}g and h when the field was applied in the $ab$ plane.
As an easy-axis ferromagnet, LaCrGe$_3$ displays a field-induced spin reorientation and a saturated magnetization of $M_{\mathrm{sat}}$ = 1.1 $\mu_{\mathrm{B}}/f.u.$. On the other hand, CeCrGe$_3$, an easy-plane ferromagnet, has $M_{\mathrm{sat}}$ = 1.3 $\mu_{\mathrm{B}}/f.u.$. 
The difference in their $M_{\mathrm{sat}}$ is smaller than that of a Kramer's doublet for a Ce atom in a hexagonal coordination, of which the smallest moment is 0.43 $\mu_{\mathrm{B}}$ determined by the crystal-electric-field ground state.
This observation unveils the screening of the Ce 4f local moments at low temperatures, in line with the HF characters.

In contrast, the Hall resistivity of CeCrGe$_3$ is two orders of magnitude greater than that of LaCrGe$_3$ at 2 K. In FM conductors, the Hall resistivity can be divided into normal and anomalous contributions, as $\rho_{\mathrm{H}} = \rho^{\mathrm{O}}+\rho^{\mathrm{A}}= R_{\mathrm{H}}B + 4\pi R_{\mathrm{s}} M$, where $R_{\mathrm{H}}$ and $R_{\mathrm{s}}$ are ordinary and anomalous Hall coefficients, respectively, $B$ is the magnetic field, and $M$ is the magnetization. 
The field dependences of $M$ and $\rho_{xz}$ in CeCrGe$_3$ at 2~K are nearly identical, indicating that the anomalous contribution dominates the Hall response, yielding $R_{\mathrm{H}} = -2.2 \times10^{-9}$ m$^3$ C$^{-1}$ and $R_{\mathrm{s}} = 4.2 \times10^{-6}$ m$^3$ C$^{-1}$. 
In contrast, the small signal of $\rho_{xz}$ in LaCrGe$_3$ shows a complicated dependence on the field, which may indicate the presence of an additional spin orientation transition in low field.
We estimated the effective carrier density $n$ as $-2.8\times10^{21}$ and  $-3.4\times10^{21}\ \mathrm{cm}^{-3}$ from the $R_{\mathrm{H}}$ values for  CeCrGe$_3$ and  LaCrGe$_3$ at 2~K, respectively (See more details in the Supplementary material, Fig. S9 online).

We now study the three transverse transport effects of charge and heat for CeCrGe$_3$. Figure \ref{Figure2}a-d demonstrate the field dependence of the magnetization, Hall resistivity, Nernst thermopower, and thermal Hall conductivity at various temperatures. At temperatures below 60~K, all three sets of transverse transport signals exhibit pronounced anomalous features, closely following the magnetization with a sharp switching behavior. We summarize the temperature dependence of the anomalous effects in Fig. \ref{Figure2}e-g. The anomalous Hall resistivity ($\rho^{\mathrm{A}}_{xz}$) reaches its peak value of 70 $\upmu\Omega$ cm at 2 K. Correspondingly, the anomalous Hall conductivity, $\sigma^{\mathrm{A}}_{zx} = \frac{\rho^{\mathrm{A}}_{xz}}{\rho_{xx}\rho_{zz}+(\rho^{\mathrm{A}}_{xz})^2}$, gradually increases with decreasing temperature and reaches its maximum value of 983 $\Omega^{-1}$ cm$^{-1}$  below 10 K. We note that the change of $\sigma^{\mathrm{A}}_{zx}$ is irrelevant to the resistivity change below 30 K.
Instead, it approximately follows the temperature dependence of saturated magnetization.
In comparison, the $\sigma^{\mathrm{A}}_{zx}$ value for LaCrGe$_3$ is only 130~$\Omega^{-1}$ cm$^{-1}$ at 2 K.

Figure \ref{Figure2}f shows that the anomalous Nernst coefficient $-S^{\mathrm{A}}_{zx}$ is as large as 9.5 $\upmu$V K$^{-1}$ at 35 K, a value greater than most records \cite{Ikhlas2017,Sakai2018,PhysRevMaterials.4.024202,Chen2021}. $S^{\mathrm{A}}_{zx}$ is proportional to temperature below 20~K, down to the lowest temperature of 5 K that we could measure. As a result, the anomalous Nernst angle $\tan\theta_{\mathrm{AN}} = \lvert S^{\mathrm{A}}_{zx}/S_{zz} \rvert$ is 76\% at 5 K. The anomalous thermoelectric conductivity $\alpha^{\mathrm{A}}_{zx}$, calculated as $\alpha^{\mathrm{A}}_{zx}=\sigma^{\mathrm{A}}_{zx} S_{xx} + \sigma_{zz} S^{\mathrm{A}}_{zx}$, exactly follows the temperature dependence of $S^{\mathrm{A}}_{zx}$ because the second term ($\sigma_{zz} S^{\mathrm{A}}_{zx}$) overshadows the expression. $-\alpha^{\mathrm{A}}_{zx}$ reaches its maximum value of 3.0 A K$^{-1}$ m$^{-1}$ at 35 K, which is comparable to the highest values reported in other ferromagnets \cite{USbTe2023Ran,PhysRevApplied.19.014026,PhysRevMaterials.4.024202,doi:10.1126/sciadv.abk1480,Guin2019}.
We note that the value of $\alpha^{\mathrm{A}}_{zx}/T$ for CeCrGe$_3$ at low temperature is about $-0.2~\mathrm{A~m^{-1}~K^{-2}}$, which is several times larger than those reported for Co$_2$MnGa ($0.05~\mathrm{A~m^{-1}~K^{-2}}$)~\cite{Sakai2018} and Co$_3$Sn$_2$S$_2$ ($0.05$--$0.1~\mathrm{A~m^{-1}~K^{-2}}$)~\cite{PhysRevX.9.041061}.

Figure \ref{Figure2}d shows that the anomalous thermal Hall conductivity ($\kappa^{\mathrm{A}}_{zx}$) reaches its maximum value of 81 mW K$^{-1}$ m$^{-1}$ at 40 K. This value is comparable to those recorded in topological magnets such as Mn$_3$Sn (40 mW K$^{-1}$ m$^{-1}$) \cite{PhysRevLett.119.056601} and Fe$_3$Sn$_2$ (86 mW K$^{-1}$ m$^{-1}$) \cite{PhysRevB.103.L201101} at room temperature. This giant anomalous thermal Hall effect leads to a significant anomalous thermal Hall angle of $\nabla T_x/\nabla T_z=\kappa_{zx}/\kappa_{xx}$, which reaches 2\% at 5~K (see Fig. \ref{Figure2}g).

\begin{figure}[!htp] 
	\centering 
	\includegraphics[width=1\linewidth]{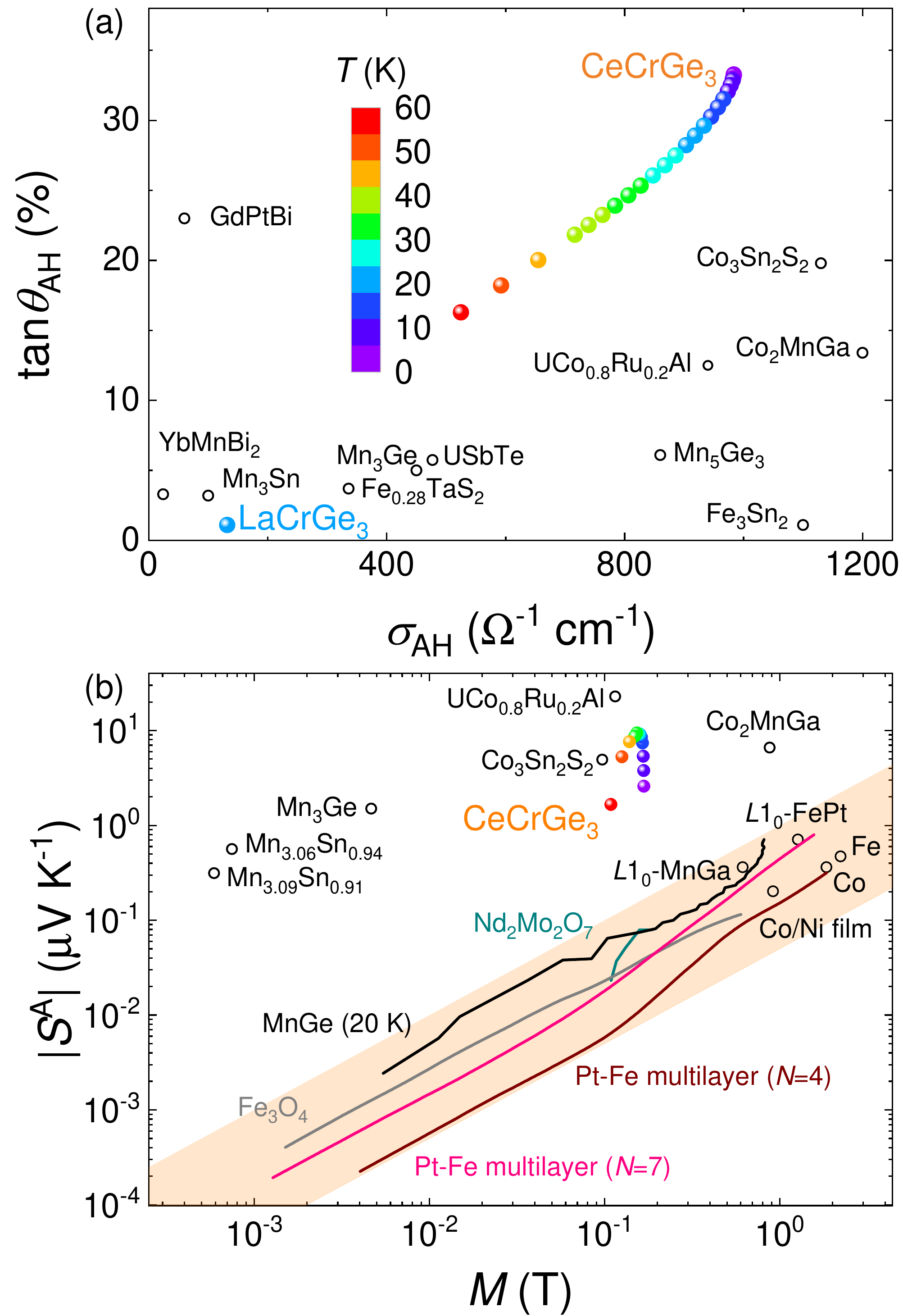} 
	\caption{Scaling of the anomalous transport coefficients. (a) The anomalous Hall angle as a function of $\sigma_{\mathrm{AH}}$ in topological magnetic materials~\cite{JDijkstra_1989,Zeng2006linear,Nakatsuji2015, doi:10.1126/sciadv.1501870, Suzuki2016,doi:10.1073/pnas.1810842115, Ye2018,Sakai2018,Liu2018,doi:10.1126/sciadv.abf1467,Pan2022,USbTe2023Ran}. 
		(b)  Anomalous Nernst coefficient ${|S^{\mathrm{A}}|}$ as a function of magnetization in magnetic materials. The ${|S^{\mathrm{A}}|}$ values for topological magnetic materials significantly surpass those for conventional magnetic materials (orange shaded region), in which ${|S^{\mathrm{A}}|}$ shows a linear dependence on the magnetization.~\cite{PhysRevLett.100.106601,PhysRevB.87.060406,PhysRevB.88.064409,Ramos2014Fe3O4,Hasegawa2015Fe,PhysRevB.92.094414,
Ikhlas2017,Sakai2018,PhysRevMaterials.4.024202,Chen2021,doi:10.1126/sciadv.abf1467}
	}
	\label{Figure3} 
\end{figure}

To evaluate the magnitude of the AHE of CeCrGe$_3$, we compare the values of $\sigma^{\mathrm{A}}_{zx}$ and  $\tan \theta_{\mathrm{AH}}$ of CeCrGe$_3$ at various temperatures, with the maximum values of other ferromagnets, including various topological magnets~\cite{JDijkstra_1989,Zeng2006linear,Nakatsuji2015, doi:10.1126/sciadv.1501870, Suzuki2016,doi:10.1073/pnas.1810842115, Ye2018,Sakai2018,Liu2018,doi:10.1126/sciadv.abf1467,Pan2022,USbTe2023Ran}, as shown in Fig. \ref{Figure3}a.
The anomalous Hall angle, $\tan \theta_{\mathrm{AH}} = \lvert {\sigma_{\mathrm{AH}}}/{\sigma} \rvert$ measures the relative contribution of the anomalous Hall current with respect to the longitudinal current.
Its magnitude serves as a key indicator of topological transport properties, as it quantifies the intrinsic efficiency of BC-driven transport in magnetic materials.
Traditional magnets such as elemental iron have a value of $\sigma_{\mathrm{AH}}$ greater than  $10^4 ~\Omega^{-1}$ cm$^{-1}$ due to the extrinsic mechanism of skew scattering but the magnitude of its $\tan \theta_{\mathrm{AH}}$ is in the order of 1\% \cite{RevModPhys.82.1539}.
The value of $\tan \theta_{\mathrm{AH}} = \lvert \sigma_{zx}^{\textrm{A}}/\sigma_{zz} \rvert$ for CeCrGe$_3$ monotonically increases with decreasing temperature, reaching 33\% at 2 K, unveiling an intrinsic mechanism underneath. 
This value is magnificent compared with the value of the non-4f analog LaCrGe$_3$ ($1.5$\%), and the values for other topological magnets.

For the ANE, we demonstrate the logarithmic plot of the values of $ \lvert S^{\mathrm{A}} \rvert$ versus magnetization of CeCrGe$_3$ at various temperatures with the maximum values of other ferromagnets in Fig. \ref{Figure3}b~\cite{PhysRevLett.100.106601,PhysRevB.87.060406,PhysRevB.88.064409,Ramos2014Fe3O4,Hasegawa2015Fe,PhysRevB.92.094414,
Ikhlas2017,Sakai2018,PhysRevMaterials.4.024202,Chen2021,doi:10.1126/sciadv.abf1467}.
The $ \lvert S^{\mathrm{A}} \rvert$ values for CeCrGe$_3$ are enhanced compared to the empirical scaling relation with $M$ as shown in the shaded region, similar to other topological magnets such as Co$_3$Sn$_2$S$_2$ \cite{PhysRevX.9.041061} and Co$_2$MnGa \cite{PhysRevB.101.180404}.
On the other hand, CeCrGe$_3$ and UCo$_{0.8}$Ru$_{0.2}$Al are highlighted as their exceptional f-electron magnetism and the $ \lvert S^{\mathrm{A}} \rvert$ values in the order of 10 $\upmu$V K$^{-1}$.

\section{Calculation}

\begin{figure}[!htp] 
	\includegraphics[width=1\linewidth]{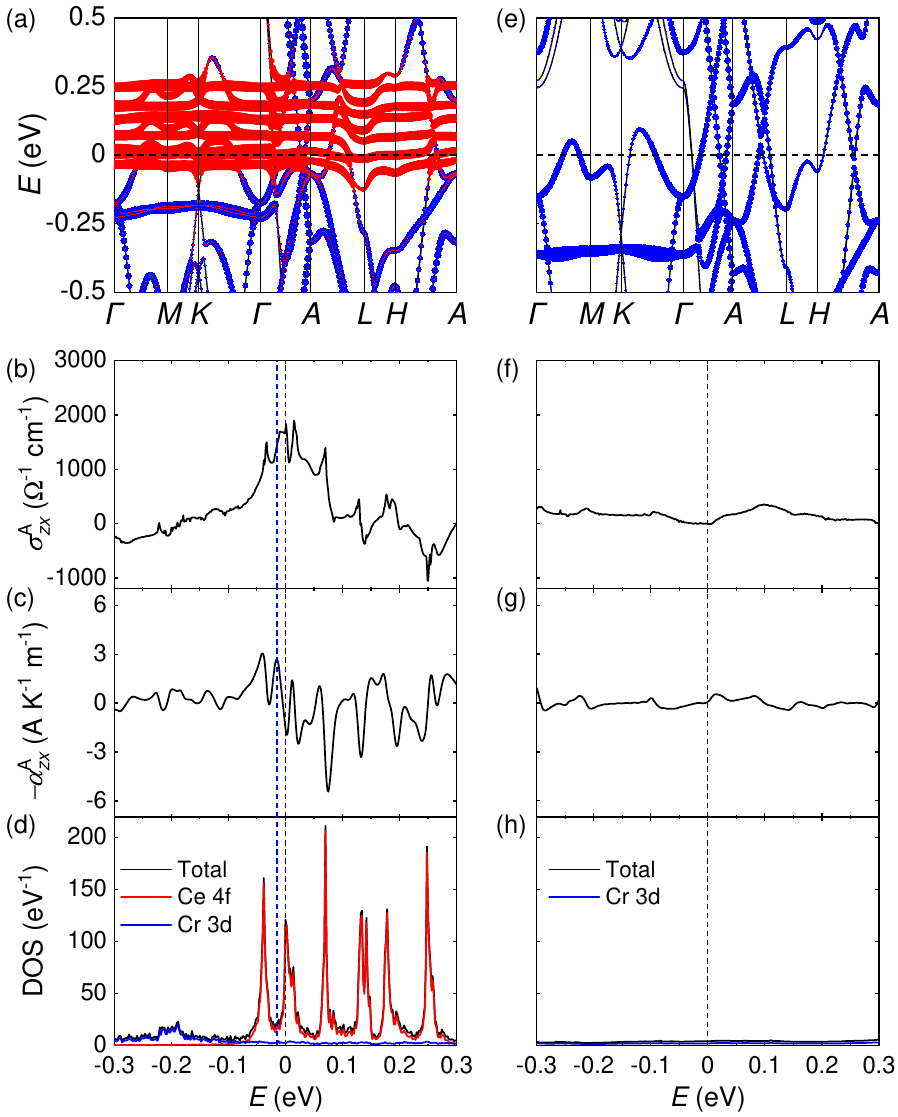} 
	\caption{DFT calculated electronic structures, $\sigma^{\mathrm{A}}_{zx}$, and $-\alpha_{zx}^{\mathrm{A}}$ for CeCrGe$_3$ and LaCrGe$_3$.
		(a) and (e) Band structures of CeCrGe$_3$ and LaCrGe$_3$, respectively. The red and blue dots represent Ce-f and Cr-d orbital projections. 
		(b)-(d) and (f)-(h) Chemical potential dependency of $\sigma^{\mathrm{A}}_{zx}$ at 0~K, $-\alpha^{\mathrm{A}}_{zx}$ at 35 K, and total and partial DOS for CeCrGe$_3$ and LaCrGe$_3$, respectively.}
	\label{Figure4} 
\end{figure}

The analyses above reveal that the large anomalous effects are the result of the BC of the electrons in CeCrGe$_3$. To understand how the 4f electrons of Ce and 3d electrons of Cr play the role in the BC, we compare the calculated results for CeCrGe$_3$ (Fig. \ref{Figure4}a-d) and LaCrGe$_3$ (Fig. \ref{Figure4}e-h) obtained from DFT in the LDA \cite{PhysRevB.23.5048}. Figure \ref{Figure4}a shows that the 4f orbitals in CeCrGe$_3$ contribute to the electronic bands with small energy dispersion in a narrow energy window near the $E_{\mathrm{F}}$. 
These flat bands give rise to a series of high-DOS peaks as demonstrated in Fig. \ref{Figure4}d.
In contrast, the DOS of LaCrGe$_3$, which is dominated by the Cr 3d orbitals, is much lower.
Of particular interest, a flat band lies near the Fermi level in CeCrGe$_3$, resulting in a high DOS, $N(E_{\mathrm{F}}) = 121~\mathrm{eV}^{-1}$. This yields an estimated electronic specific heat coefficient of $\gamma = 285~\mathrm{mJ\,mol^{-1}\,K^{-2}}$, in good agreement with experimental values, highlighting the strong electronic correlations. In contrast, the value of $N(E_{\mathrm{F}})$ is $3.9~\mathrm{eV}^{-1}$ for LaCrGe$_3$.

Our calculation also shows that LaCrGe$_3$ and CeCrGe$_3$ have 4 and 44 pairs of Weyl points, respectively, in an energy range of $\pm$ 20 meV near the $E_{\mathrm{F}}$ (refer to the Supplementary material, Fig. S12 online).
As shown in Fig.~\ref{Figure4}a, the orbital-projected band structures reveal that the Weyl nodes in CeCrGe$_3$ mainly originate from band crossings between hybridized Ce 4f and Cr 3d states near the Fermi level. The strong correlation of the 4f electrons leads to quasi-particle renormalization, reducing the bandwidth and bringing multiple band crossings into close proximity to the Fermi level, while the f--d hybridization enhances inter-orbital mixing. These effects together amplify the BC around the hybridized bands, resulting in enhanced anomalous Hall and Nernst responses. In contrast, LaCrGe$_3$, which lacks partially filled 4f orbitals, shows significantly fewer Weyl nodes and much weaker BC near the Fermi level (Fig.~\ref{Figure4}e).
We note that the calculated values of $-\alpha_{zx}^{\mathrm{A}}$ and $\sigma^{\mathrm{A}}_{zx}$ are highly sensitive to a delicate change of $E_{\mathrm{F}}$.
As shown in Fig. \ref{Figure4}b and c, when we set the energy at $E$ = $E_{\mathrm{F}} -$15 meV, we obtain $\sigma_{zx}^A$ being about $ 1400~\Omega^{-1}\,\text{cm}^{-1}$ at 0~K and $-\alpha_{zx}^A$ being about $2.7~\text{A}\,\text{m}^{-1}\,\text{K}^{-1}$ at 35~K, which are in rough agreement with the experimental values.
As comparison, the calculated $\sigma_{zx}^{\mathrm{A}}$ and $-\alpha_{zx}^{\mathrm{A}}$ values for LaCrGe$_3$ are close to zero in an extensive energy range, as shown in Fig. \ref{Figure4}f and g, respectively.

\section{Discussion} 

\begin{figure}[!htp] 
	\includegraphics[width=1\linewidth]{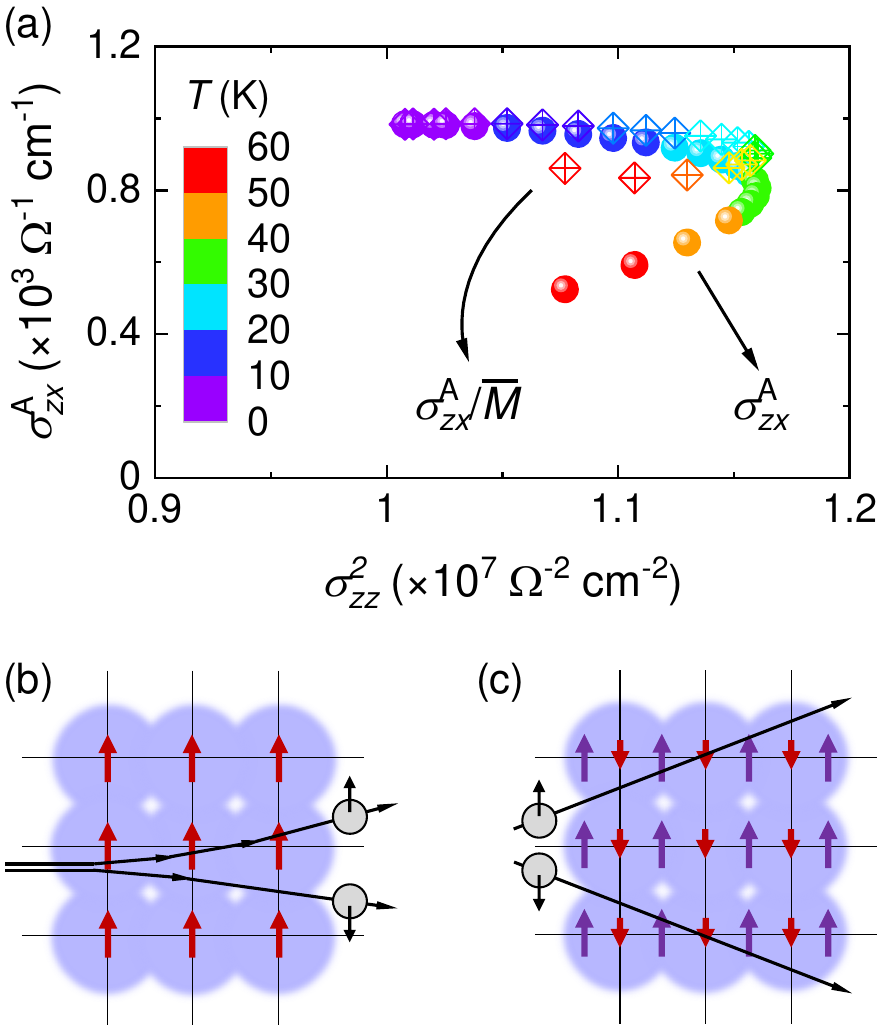} 
	\caption{Scaling of the AHE in CeCrGe$_3$ and schematic illustration of intrinsic and extrinsic AHE in f-electron compounds.
    (a) Scaling relation of ${\sigma^{\mathrm{A}}_{zx}}$ as a function of $\sigma^{2}_{zz}$ at various temperatures. The values of ${\sigma^{\mathrm{A}}_{zx}}/{\overline{M}}$ are also presented for comparison, where the magnetization $\overline{M}$ is normalized as $\overline{M}={M_{\mathrm{sat}}(T)}/{M_{\mathrm{sat}}}(T=2\mathrm{\ K})$. 
    (b) Sketches of the extrinsic skew-scattering AHE due to the Kondo moments such as in CeTiGe$_3$, in which only f-moment presents (red arrows). (c) The intrinsic AHE due the BC in topological flat band such as in CeCrGe$_3$, in which the 3d-moments (blue arrows) are anti-parallel with the f-moments (red arrows). }
	\label{Figure5} 
\end{figure}

The existence of a flat band near $E_{\mathrm{F}}$ in CeCrGe$_3$ gives rise to a heavy electron mass ($m^*$) and a low Fermi temperature ($T_{\mathrm{F}}$), which are the signatures of HF.
According to our measurements of the $\gamma$ and $n$ values at low temperatures, we estimate $T_{\mathrm{F}}$ by using the equation  $T_{\mathrm{F}}={\pi^2 k_{\mathrm{B}} n}/{2\gamma}$, where $k_{\mathrm{B}}$ is Boltzmann's constant \cite{10.1093/acprof:oso/9780199697663.001.0001}.
The result is $T_{\mathrm{F}}= 1040$ and 100~K for LaCrGe$_3$ and CeCrGe$_3$, respectively.
Moreover, by assuming a spherical Fermi surface, $\gamma = \frac{1}{3}k^2_{\mathrm{B}}k_{\mathrm{F}}m^*/\hbar^2$ and $E_{\mathrm{F}}={\hbar^2k^2_{\mathrm{F}}}/{2m^*}$, we estimate the effective electron mass as $m^*=(\frac{9\hbar^6 \gamma^2}{2k^5_{\mathrm{B}}T_{\mathrm{F}}})^{\frac{1}{3}}= 83\ m_e$ for CeCrGe$_3$.
In contrast, the value of $m^*$ is only $9.3\ m_e$ for LaCrGe$_3$.

Comparing the values of $m^*$ and  $T_{\mathrm{F}}$ for CeCrGe$_3$ and LaCrGe$_3$, we highlight the role of the 4f electron and the Kondo effect on the flat band near the Fermi level.
As a result, the electric transport in CeCrGe$_3$ is characterized by a large Fermi surface, which is distinct from those of semimetals, although they all have significantly low  $E_{\mathrm{F}}$.
The topological semimetals, such as GdPtBi \cite{doi:10.1073/pnas.1810842115} and Co$_3$Sn$_2$S$_2$ \cite{PhysRevX.9.041061}, show significant AHE due to their Fermi level close to the Weyl points.
Their electric transport is characterized by highly mobile carriers whose density is about $n=10^{19}$ cm$^{-3}$, and this value is much smaller than that of CeCrGe$_3$ which is on the order of $10^{21}~\mathrm{cm^{-3}}$.
   Another character of the electric transport of CeCrGe$_3$ is the large residual resistivity, which originates from the Kondo-hybridized bands and can be regarded as a flat-band effect~\cite{Ma2025anisotropic}. 
The exceptional $\tan \theta_{\mathrm{AH}}$ value of CeCrGe$_3$ can be attributed to the large residual resistivity and significant intrinsic AHE.

To shed light on the relationship of the AHE and the Kondo effect in CeCrGe$_3$, we conduct a scaling analysis of $\sigma^{\mathrm{A}}_{zx}$ and $\sigma_{zz}^2$ in Fig.~\ref{Figure5}a.
Generally, the AHE is caused by an extrinsic skew scattering resulting in a quadratic relationship between $\sigma^{\mathrm{A}}_{zx}$ and $\sigma_{zz}$  ($\sigma^{\mathrm{A}}_{zx}\sim \sigma_{zz}^2$)~\cite{tian2009proper}, or an intrinsic BC causing  $\sigma^{\mathrm{A}}_{zx}$ to be independent of $\sigma_{zz}$~\cite{BerryXiao2006,Onoda2008quantum}.
In general, $\sigma_{\mathrm{AH}}$ and $\sigma^2$ exhibit an approximately linear relation, which allows the intrinsic and extrinsic contributions to be separated through scaling analysis.

However, in CeCrGe$_3$, we observe a ``$\supset$-shaped'' dependence with a knee point at about 30~K, indicating that a single scaling law cannot describe the data below $T_{\mathrm{C}}$. Considering that the temperature dependence of the magnetization may affect the value of $\sigma^{\mathrm{A}}_{zx}$ due to spin fluctuation \cite{Zeng2006linear}, we further take the temperature dependence of the magnetization into account, as shown in Fig.~\ref{Figure5}a. Even in this case, the shape of $\sigma^{\mathrm{A}}_{zx}/\overline{M}$ remains essentially unchanged. A similar phenomenon has also been reported in the 5f compound USbTe~\cite{USbTe2023Ran}, where a crossover from extrinsic AHE at high temperatures to intrinsic AHE at low temperatures was observed.

Unlike USbTe~\cite{USbTe2023Ran}, in the case of CeCrGe$_3$, the value of $\tan\theta_{\mathrm{AH}}$ is greater than 15\% event at 60~K, far exceeding the expectation from skew-scattering mechanisms.
This suggests that the intrinsic contribution remains dominant in the high-temperature regime.
We suggest that the break of the universal scaling relation is due to the fact that the BC fades out at elevated temperatures as long as the temperature dependent magnetization is taken into account.
Notably, the Kondo coherence temperature $T_{\mathrm{coh}}$ is estimated as 50~K, about which the Kondo-derived flat band should gradually break down.
This conspicuous breakdown of the AHE scaling relation is another piece of evidence of the Kondo-flat-band-driven BC in CeCrGe$_3$.

We note that the observed giant AHE distinguish CeCrGe$_3$ from the ordinary HF compounds in which the resonant scattering plays a vital role in their electric transport  \cite{PhysRevLett.55.414,PhysRevB.36.1907,doi:10.1143/JPSJ.63.2627,doi:10.1080/00018732.2012.730223,PhysRevB.87.045102}.
An ideal comparison of CeCrGe$_3$ is iso-structural, HF ferromagnet CeTiGe$_3$, which has $T_{\mathrm{C}}$ being 14~K and $\gamma $ being 75~$\ \mathrm{mJ\ mol^{-1}\ K^{-2}}$  \cite{Li2025Extrinsic}.
With only Ce 4f magnetism, the AHE in CeTiGe$_3$ originates from resonant skew scattering and the value of $\tan \theta_{\mathrm{AH}}$ is limited to about 1--2\%~\cite{Li2025Extrinsic}.
Apparently, the critical difference lies in the existence of the topological Kondo flat band.
In CeTiGe$_3$ and other ordinary HF systems, the conduction electron is non-magnetic, and the magnetic ordering is formed due to the Ruderman–Kittel–Kasuya–Yosida (RKKY) interaction between the partially screened local moments.
In this case, the local moments serve as effective scattering centers for skew scattering, resulting in a conventional extrinsic AHE, as shown in Fig.~\ref{Figure5}b.
In contrast, the hybridization between the magnetic d electrons and f electrons effectively quenches the local moments, leading to the formation of topological Kondo flat bands that host significant BC and thus give rise to the giant intrinsic anomalous effects, as illustrated in Fig.~\ref{Figure5}c.

A large intrinsic AHE does not guarantee a large ANE because the intrinsic AHE is determined by the summation of the BC for all the occupied states, while the ANE is given by the BC at the $E_{\mathrm{F}}$ \cite{RevModPhys.82.1959}.
To understand how the flat band plays the role in the significant ANE, we rewrite the Mott relation as ~\cite{crossoverbehaviorTukura2007,Lee2004CuCr2Se4}
\begin{eqnarray}
	&  &\frac{\alpha^{\mathrm{A}}}{T}\Big|_{T\rightarrow 0} \\ \notag
    & = & -\frac{\pi^2}{3} \frac{k^2_{\mathrm{B}}}{e}\frac{\mathrm{d} n}{\mathrm{d}E} \frac{\mathrm{d}\sigma^{\mathrm{A}}(E)}{\mathrm{d}n}\Big|_{E=E_{\mathrm{F}}} \\\notag
	& = & -\frac{\pi^2}{3} \frac{k^2_{\mathrm{B}}}{e}\rho({E_{\mathrm{F}}})\frac{-\frac{e^2}{\hbar}\int_{E_{\mathrm{F}}}^{E_{\mathrm{F}}+\mathrm{d}E}\frac{\mathrm{d}\boldsymbol{k}}{(2\pi)^3}\Omega_{xy}(\boldsymbol{k})}{\int_{E_{\mathrm{F}}}^{E_{\mathrm{F}}+\mathrm{d}E}\frac{\mathrm{d}\boldsymbol{k}}{(2\pi)^3}} \\ \notag
	& = & -\frac{\pi^2}{3} \frac{k^2_{\mathrm{B}}}{e}\frac{e^2}{\hbar}\rho({E_{\mathrm{F}}})\langle \Omega(E_{\mathrm{F}}) \rangle.
\end{eqnarray}
Here $\langle \Omega(E_{\mathrm{F}}) \rangle$ represents the average BC of the states at the Fermi level because $\sigma^{\mathrm{A}}$ represents the change in the integrated BC of occupied states per unit energy interval, while $\mathrm{d}n$ corresponds to the change in the number of occupied states within the same interval.  
With a substantial BC near $E_{\mathrm{F}}$, a large DOS at the Fermi level is crucial for achieving an enhanced value of $\alpha^{\mathrm{A}}/T$, which is vividly manifested in our DFT calculations. As shown in Fig.~\ref{Figure4}a, within the energy window from $E_{\mathrm{F}}-0.1$~eV to $E_{\mathrm{F}}+0.3$~eV, six weakly dispersive flat bands dominated by f orbitals are observed. These flat bands give rise to six pronounced and sharp peaks in the DOS, as displayed in Fig.~\ref{Figure4}d, each of which corresponds to a prominent peak in $\alpha^{\mathrm{A}}_{zx}$ as shown in Fig.~\ref{Figure4}c.

From an entropic perspective,  it is well known that the Seebeck coefficient represents the ratio of entropy flow to charge flow and therefore it is on the order of $k_{\mathrm{B}}/e$, which also sets the upper bound for the ratio of anomalous entropy to charge.
For a broad class of topological magnets, the ratio $\alpha^{\mathrm{A}}/\sigma^{\mathrm{A}}$ increases monotonically with increasing temperature and approaches the value of $k_{\mathrm{B}}/e$ near $T_{\mathrm{C}}$, where the anomalous Nernst coefficient $S^{\mathrm{A}}$ reaches its maximum~\cite{crossoverbehaviorTukura2007,Pu2008GaMnAs,Mn3GeXu2020,PhysRevB.101.180404}.
UCo$_{0.8}$Ru$_{0.2}$Al and YbMnBi$_2$, which demonstrate giant ANE, have exceptional ratio of  $\alpha^{\mathrm{A}}/\sigma^{\mathrm{A}}$ at finite temperature due to multi-band effects~\cite{doi:10.1126/sciadv.abf1467,Guo2023Onsager}.

\begin{figure}[!htp] 
	\centering 
	\includegraphics[width=1\linewidth]{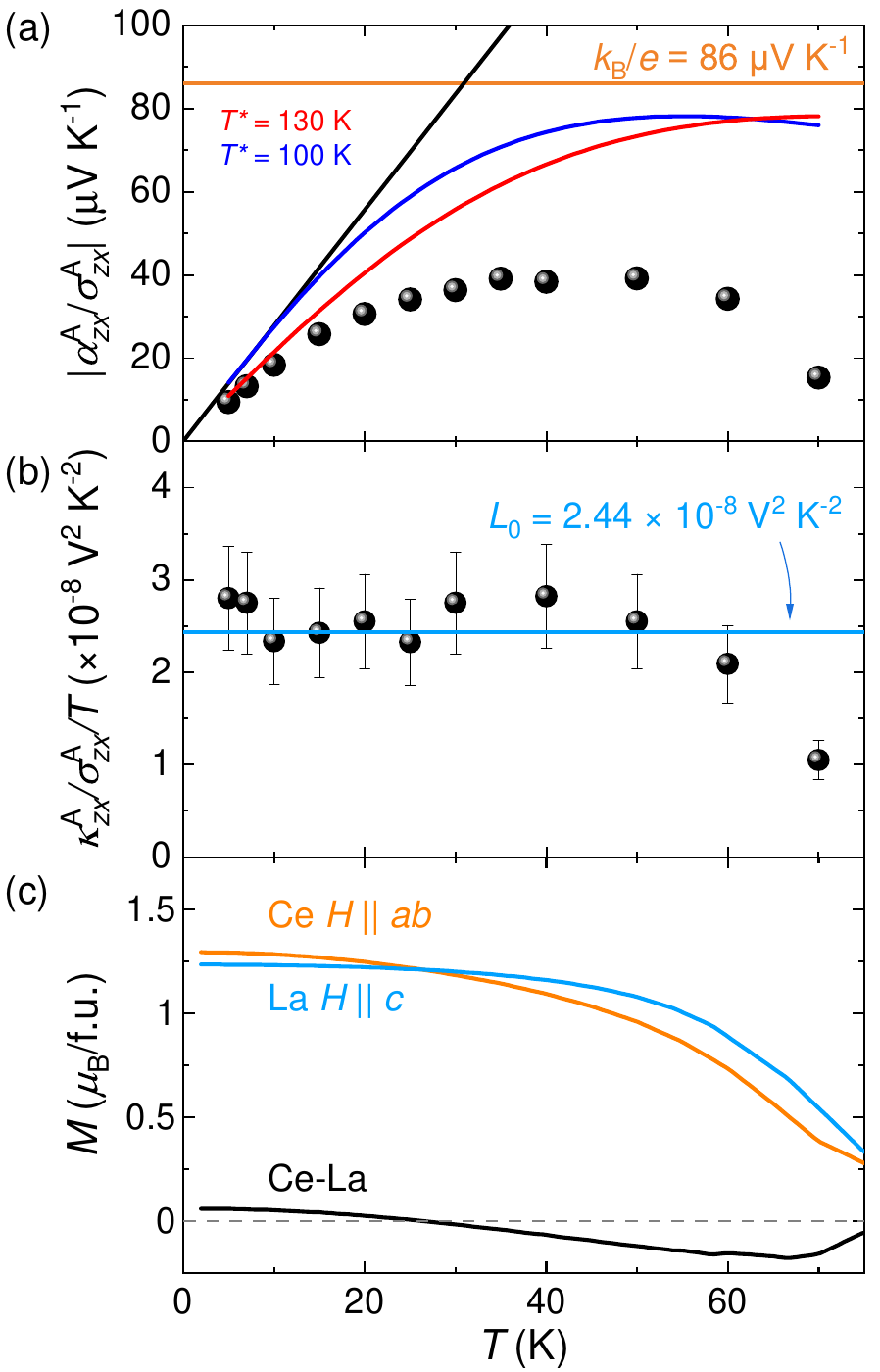} 
	\caption{Anomalous Mott relation and anomalous Wiedemann–Franz law in CeCrGe$_3$ (a) ${|\alpha^{\mathrm{A}}_{zx}}/{\sigma^{\mathrm{A}}_{zx}}|$ as a function of $T$ for CeCrGe$_3$. The orange line indicates $k_{\mathrm{B}}/e$, and the slope of the black line is $2\gamma/3en$. The red and blue curves represent the fitting results of Eq. \ref{eq5} when $T^{\ast}$ is 130 and 100 K, respectively.  
		(b) ${\kappa^{\mathrm{A}}_{zx}}/{\sigma^{\mathrm{A}}_{zx}}/T$ for CeCrGe$_3$. The blue line represents the value of $L_{0}$. The error bar is estimated as $\pm$20\% considering the uncertainty in the transport measurements and the samples' geometric factor. (c) The temperature dependence of the saturated magnetic moment for CeCrGe$_3$, LaCrGe$_3$, and their difference. For comparison, the temperature scale of LaCrGe$_3$ is normalized as $T/T_{\mathrm{C}}(\mathrm{La}) \times T_{\mathrm{C}}(\mathrm{Ce})$, where $T_{\mathrm{C}}(\mathrm{La})$ and $T_{\mathrm{C}}(\mathrm{Ce})$ denote the Curie temperatures of LaCrGe$_3$ and CeCrGe$_3$, respectively.
}
	\label{Figure6} 
\end{figure}

As shown in Fig.~\ref{Figure6}a, unlike other topological magnets, the ratio $\alpha^{\mathrm{A}}_{zx}/\sigma^{\mathrm{A}}_{zx}$ in CeCrGe$_3$ only approaches the linear relation in the zero-temperature limit and quickly reaches a maximum value of about $k_{\mathrm{B}}/2e$ at 35~K, which is unexpectedly low given the large $S^{\mathrm{A}}$ in CeCrGe$_3$.
To understand this nonlinear behavior of the Mott relation, we perform the Sommerfeld expansion, with the help of the Dirac model, to modify the Mott relation as~\cite{Qiang2023Topological}:
\begin{equation}
	\frac{\alpha^{\mathrm{A}}}{\sigma^{\mathrm{A}}}=\frac{eL_0T}{\mu[1+(\pi^2/3)(k_{\mathrm{B}}T/\mu)^2]},
	\label{eq5}
\end{equation}
where $L_0$ is the Lorenz constant.
This equation provides an analytical formula only dependent on the chemical potential $\mu$ and has been proven to be able to explain the nonlinear Mott relation and the violation of Wiedemann–Franz law (WFL) for TbMn$_6$Sn$_6$ and Mn$_3$Ge~\cite{Qiang2023Topological}.

In the limit of $ k_{\mathrm{B}}T \ll \mu$, the second term of the denominator in Eq. \ref{eq5} is negligible, and the ratio exhibits a linear temperature dependence.
Since $\mu$ can be expressed in terms of the carrier density $n$ and the electronic specific heat coefficient $\gamma$, Eq.~\ref{eq5} can be rewritten as
\(
\left|{\alpha^{\mathrm{A}}_{xy}}/{\sigma^{\mathrm{A}}_{xy}}\right|
= {2\gamma T}/{3 e n} ,
\)
which corresponds to the black solid line shown in Fig.~\ref{Figure6}a.
When $k_{\mathrm{B}}T$ becomes comparable to $\mu$, Eq.~\ref{eq5} predicts that the ratio exhibits a nonlinear temperature dependence and reaches an upper bound of approximately $0.9\,k_{\mathrm{B}}/e$ at $T = \sqrt{3}\mu/\pi k_{\mathrm{B}}$.
For CeCrGe$_3$, we observe a pronounced nonlinear behavior of $\alpha^{\mathrm{A}}_{xy}/\sigma^{\mathrm{A}}_{xy}$ emerging around $\sim 20$~K, which reaches a maximum at approximately 40~K.
The best fitting result with Eq. \ref{eq5} is shown by the red curve in Fig.~\ref{Figure6}a, leads to a $\mu$ value corresponding to $T^{\ast}=\mu/k_{\mathrm{B}}=130$~K which is close to $T_{\mathrm{F}}=100$~K.
However, this model, implying a low $T_{\mathrm{F}}$, still fails to fully capture the pronounced nonlinear behavior of the Mott relation observed in CeCrGe$_3$.

Then we check the validity of the anomalous WFL, $\kappa^{\mathrm{A}}_{zx}/\sigma^{\mathrm{A}}_{zx} = L_0T$, by plotting the ratio of $\kappa^{\mathrm{A}}_{zx}/\sigma^{\mathrm{A}}_{zx}/T$ versus $T$, as shown in Fig. \ref{Figure6}b.
The validity of the anomalous WFL below 60 K indicates that the anomalous transport 
in this temperature range is dominated by elastic scattering processes, which 
further proves that the breakdown of the AHE scaling relation is not due to 
the emergence of extrinsic contribution at elevated temperatures.
Instead, this breakdown reflects a pronounced evolution of the topological band structure at finite temperatures due to the Kondo effect and strong f–d hybridization.
This evolution gives rise to two consequences that affect the ANE significantly and lead to the non-linear Mott relation.
Firstly, the BC is greatly enhanced at low temperatures when the Kondo flat band is condensed.
Secondly, because the quasiparticle spectrum of the Kondo flat band is renormalized, the reduced energy scale naturally enhances the sensitivity of thermoelectric responses to temperature and invalidates the applicability of the Sommerfeld expansion. 
Therefore the linear anomalous Mott relation only takes place at the low-temperature limit in the HF magnets.

Although a coherence peak is absent in the $\rho_{zz}(T)$ curve of CeCrGe$_3$, the condensation of the Kondo flat band can be glimpsed when we compare the temperature dependence of the saturated magnetization $M_{\mathrm{sat}}(T)$ in CeCrGe$_3$ and LaCrGe$_3$, as shown in Fig.~\ref{Figure6}c. 
The saturated magnetization of CeCrGe$_3$ is less than that of LaCrGe$_3$ by approximately $0.2~\mu_{\mathrm{B}}$ per formula unit near $T_{\mathrm{C}}$. 
This difference gradually diminishes and becomes nearly zero upon cooling, which is due to the progressive screening of the Ce magnetic moments.
As a consequence, the f–d hybridization builds up the Kondo flat band with strong BC at the Fermi level.

Recent progress has highlighted a pathway toward understanding correlation-driven topology within HF systems~\cite{Paschen2021topoKondo, Checkelsky2024flatband}. 
A Weyl-Kondo semimetal state with strongly renormalized Weyl nodes can manifest enhanced BC and a unique spontaneous Hall effect at ultra-low temperature~\cite{doi:10.1073/pnas.1715851115, WeylKS2019, Chen2022topocorre}.
On the other hand, significant anomalous effects have been observed in several U-based HF compounds~\cite{USbTe2023Ran,doi:10.1126/sciadv.abf1467,Xu2024tunable}.
The observation of significant intrinsic anomalous effect in CeCrGe$_3$ and CeCo$_2$As$_2$ ~\cite{Guan2025CeCo2As2} points a new route for searching for topological phenomena in the HF magnets.
As ordinary Ce-based intermetallics usually demonstrate low-temperature antiferromagnetic ordering, the compounds containing both Ce atoms and magnetic transition-metal elements can simultaneously host strong magnetism with high magnetic transition temperatures, as well as pronounced HF characteristics associated with the Ce 4f electrons. 
We believe this combination is particularly favorable for exploring topological properties, and some theoretical understanding has been performed ~\cite{Hu2025CeCo2P2}.
Moreover, it has been noticed that topological magnets have the potential to be used for constructing transverse thermoelectric devices, and we believe the HF magnets have the kind of potential working in a cryogenic environment.

\section*{Conflict of interest}
The authors declare that they have no conflict of interest.

\section*{Acknowledgments}
This work was supported by the National Natural Science Foundation of China (12225401, 12141002, 12274154, and 12404182), the National Key Research and Development Program of China (2021YFA1401902 and 2024YFA1611200), Quantum Science and Technology --- National Science and Technology Major Project (2021ZD0302600), and the Interdisciplinary program of Wuhan National High Magnetic Field Center (WHMFC2025003), Huazhong University of Science and Technology.
The computation is completed in the HPC Platform of Huazhong University of Science and Technology.
We thank Xitong Xu, Yan Sun, and Zijia Cheng for helpful discussions.

\section*{Author contributions}
Under the supervision of Shuang Jia, the samples were  synthesized and experimental measurements were performed by Longfei Li, Shuyue Guan, Jiawei Li, and Xinxuan Lin. Under the supervision of Gang Xu, the DFT+DMFT calculations and DFT calculations were carried out by Shengwei Chi and Jianzhou Zhao. The manuscript was written by Shuang Jia, Shuyue Guan, Longfei Li, Shengwei Chi, and Gang Xu.

\bibliographystyle{scibull-elsarticle-num}
\bibliography{CeCrGe3-mod}

\end{document}